\def\spose#1{\hbox to 0pt{#1\hss}}
\def\simlt{\mathrel{\spose{\lower 3pt\hbox{$\mathchar''218$}}
     \raise 2.0pt\hbox{$\mathchar''13C$}}}
\def\simgt{\mathrel{\spose{\lower 3pt\hbox{$\mathchar''218$}}
     \raise 2.0pt\hbox{$\mathchar''13E$}}}
\documentstyle[aaspp4,epsf]{article} % this is for small print
\begin{document}
 
%macro for invoking today's date (when TeX is run on your file)
\def\today{\number\year\space \ifcase\month\or  January\or February\or
        March\or April\or May\or June\or July\or August\or
        September\or
        October\or November\or December\fi\space \number\day}
%\fraction makes a nice fraction
\def\fraction#1/#2{\leavevmode\kern.1em
 \raise.5ex\hbox{\the\scriptfont0 #1}\kern-.1em
 /\kern-.15em\lower.25ex\hbox{\the\scriptfont0 #2}}
\def\heion{\ion{He}{2}}

\title{Theoretical Transmission Spectra During Extrasolar Giant Planet Transits}
\author{S. Seager\footnote{School of Natural Sciences, Institute for
Advanced Study, Olden Lane, Princeton, NJ, 08540},
D. D.~Sasselov\footnote{Astronomy Department, Harvard University, 60 Garden
St., Cambridge, MA, 02138}$^,$\footnote{Alfred P. Sloan Research Fellow}}
 
\pagestyle{plain}

\begin{abstract}
The recent transit observation of HD~209458~b --- an extrasolar planet
orbiting a sun-like star --- confirmed that it is a gas giant
and determined that its orbital inclination is 85$^{\circ}$.
This inclination makes possible
investigations of the planet atmosphere.
In this paper we discuss the planet transmission spectra
during a transit.
The basic tenet of the method is that the planet atmosphere absorption
features will be superimposed on the stellar flux as the stellar
flux passes through the planet atmosphere above the limb.
The ratio of the planet's transparent atmosphere area to the
star area is small ($\sim 10^{-3}$ -- $10^{-4}$);
for this method to work very strong planet spectral features 
are necessary. We use our models of close-in extrasolar
giant planets
to estimate promising absorption signatures: the alkali metal lines,
in particular the Na~I and K~I resonance doublets,
and the He~I $2^3S$ -- $2^3P$
triplet line at 1083.0~nm. If successful, observations will constrain
the line-of-sight temperature, pressure, and density. The most important
point is that observations will constrain the cloud depth, which in
turn will distinguish between different atmosphere models.
We also discuss the potential of this
method for EGPs at different orbital distances and orbiting
non-solar-type stars.
\end{abstract}

\section{Introduction}
The recent transit detection of HD~209458~b (Charbonneau et al. 1999a;
Henry et al. 1999ab) 
concludes a period of eager anticipation since the first
close-in extrasolar giant planet (CEGP), 51 Peg b, was discovered in 1995
(Mayor \& Queloz 1995).
Five close-in extrasolar giant planets with orbital radii $\leq 0.05$~AU
are known, and there are an additional six closer than 0.15~AU
(see e.g. Schneider 1999). The chance of a transit for a CEGP --- assuming
random alignment of the orbital inclination --- is roughly
10\%. Transits have been excluded
for five of the 11 above planets (Henry et al. 2000, 1997;
Baliunas et al. 1997; Henry, private communication.)
The transit of HD~209458~b confirms that the CEGPs are gas giants,
gives the planet radius, and fixes the orbital inclination, which
removes the $\sin i$ ambiguity in mass and provides the average planet
density.

While the transit gives important physical parameters
for the planet, it cannot provide
any information about the planet's atmosphere. Nevertheless, the 
near edge-on
orbital inclination means that HD~209458~b is promising for a number
of different types of planet atmosphere
investigations. Two of these involve optically reflected
light and benefit from the nearly full phase
of the planet: spectral separation of the combined star-planet light
(Cameron et al. 1999; Charbonneau et al. 1999),
and photometric observations of the phase curve
from reflected planetary light (Seager, Whitney, \& Sasselov 1999).
(See Seager et al. 1999 for a full discussion.)
A third method is the transmission spectra discussed here,
which requires the near edge-on orientation. 

The idea of spectral transmission observations is not new: they have
been observed in binary stars (e.g. Eaton 1993); they have been
extensively observed and analyzed for occultations
of stars and the Sun by planets in our Solar System
(e.g. Smith \& Hunten 1990); they are one  
motivation for large transit surveys of stars with no known CEGPs 
(e.g. Vulcan Camera Project (PI W. Borucki), STARE (PI
T. Brown), WASP (PI S. Howell));
they have been discussed briefly for extrasolar planets
(e.g. Schneider 1994; Charbonneau et al. 1999a);
and they have been discussed
for extrasolar planet exospheres (Rauer et al. 2000).
Here, for the first time
to our knowledge, we quantify the method for CEGPs and provide
estimates of specific spectral features in the combined star-planet
light during a planet transit.
Many more CEGPs will be detected in the near future, both by
ongoing radial-velocity searches and by wide-field transit searches.
The transit transmission lines
will provide us with constraints on cloud-top location, and 
line-of-sight column density, temperature ($T$), and pressure ($P$).
%along the line-of-sight through the planet's transparent atmosphere.

\section {Transmission Spectra}
\label{Observing}
During an EGP planetary transit, the planet passes in front
of the star and occults the stellar flux in the amount
equal to the ratio of the planet-to-star area.
During the transit, some of the stellar flux will pass through the
optically thin part of the planet atmosphere, the
part of the atmosphere above the planet limb.
In stellar occultations by planets in the Solar System, the limb of giant
planets is usually defined at (1) the cloud tops, or (2) at the
1-bar level (Atreya 1986). Here we define
the planet limb as the boundary (e.g. the cloud tops)
above which the planet's atmosphere is transparent to the
stellar continuum radiation. The clouds
are taken to be 1 pressure scale height above
the cloud base, which due to irradiation heating
is expected to be well above the 1 bar level (Seager 1999). 
We call the entire atmosphere above
the limb the ``transparent atmosphere'', although the transparent
atmosphere is optically thick in some transitions.
Below the limb the optically thick clouds
prevent radiation from being transmitted through the
atmosphere.

The ratio of the
planet's projected transparent atmosphere area to star-minus-planet
area is small, on the order of $10^{-4}$ -- $10^{-3}$, using
$R_*=1.3R_{\odot}$, $R_P=1.54 R_J$ (based on HD~209458 parameters
from Mazeh et al. 2000),
and using an estimate for the
limb radial depth of 0.01~$R_P$ to 0.05~$R_P$ .
The planet's
absorption features will be superimposed on the observable stellar flux,
and will appear at the
$\sim 10^{-4}$ level below the continuum flux.
Very strong spectral features are needed for detection;
essentially the absorption
features must be optically thick.
In contrast to reflected planetary light, the transit transmission
flux is not diluted by the planet-star distance,
and is not cloud albedo-dependent.

During Solar System planet occultations and binary star
occultations successive measurements are made as the planet occults
the star (i.e. during ingress or egress). The time-dependent
change in spectral features provide column density and temperatures
for different atmosphere heights. Because the CEGP is much smaller
and fainter than the parent star this will not be possible
in the near future.
The best orbital phase for the transmission spectra observations
of CEGPs is when the planet is fully projected on the visible hemisphere
of the star so that the planet's projected transparent atmosphere takes out
the greatest area and limb darkening from the star is at its minimum.

Refraction through the CEGP's atmosphere has to be accounted for
in defining the limb and in computing the total optical path of rays
reaching us through it. During the solar transit of Venus in 1761 an 
obvious refraction effect at second contact convinced Lomonosov in
St.Petersburg that Venus had an atmosphere (Cruikshank 1983).
However, the cloud tops in our model of
HD~209458b (see below) are high in its atmosphere, where the gas
density is fairly low and refraction is small. 
In an isothermal atmosphere with gas scale height $H$, the angle of 
refraction for a ray passing at planetocentric distance $r$ is:
$\theta = \nu (r)[(2\pi R_P)/H]^{1/2}$, where $\nu (r) = n-1$
is the atmospheric refractivity at $r$ (for a $H_2+He$ mixture at STP:
$\sim 1.2\times 10^{-4}$), with $n(\lambda,r)$ the index of refraction.
Refraction introduces a lengthening of the pathways of the rays
through a spherical stratified atmosphere:
${\Delta}s = 0.5z{\theta}^2$, where simply $z^2 = (R_P+H)^2-R_P^2$.
Unlike stellar occultations in our Solar system, in CEGP transits
the parent star is a very extended background source and
subtends a significant solid angle at its orbit. This allows
us to observe rays deflected at angles larger than the average isothermal
$\theta$ through the densest optically thin layers of the atmosphere.
However, the cloud tops in HD~209458b are still at
$P< 10^5$dyn cm$^{-2}$, and ${\Delta}s/s <$ 2\%.

\subsection{Close-in EGPs Transparent Atmosphere Model}
We consider the only currently known EGP system with an observed transit,
HD~209458.
We use the stellar parameters
$R_*=1.3R_{\odot}$, T$_{\rm eff }$=6000, log $g$ = 4.25,
and [Fe/H] = 0.0,
derived from evolutionary calculations and fits to spectroscopic
data in Mazeh et al. (2000). We use the planetary parameters
$R_P=1.54~R_J$, $M_P = 0.69 M_{\rm J}$, log $g$ = 2.9,
and $i=85.2^{\circ}$ derived from the transit
observations and radial velocity observations, together with
the stellar parameters (Mazeh et al. 2000). For a circular orbit
we derive a semimajor axis $a=0.0468$~AU.
We compute the incident flux of HD~209458 with the above parameters
from the model grids of Kurucz (1992).

We compute the CEGP atmosphere (temperature-pressure (T-P) profile and
emergent spectra) using our code described in
Seager (1999) and Seager et al. (1999). 
This code is improved over the
version described in Seager \& Sasselov (1998) in two major ways.
One is a Gibbs free energy minimization code to calculate solids
and gases in chemical equilibrium; the second is condensate opacities
for 3 solid species. So while in Seager \& Sasselov (1998) we considered
neither the depletion of TiO nor formation of
MgSiO$_3$, in the new models we do. One of the largest uncertainties
in the atmosphere models is the location of clouds, and the cloud particle
type and size.
In Seager et al. (1999) we find that the T-P
profile and the emergent flux (reflected + thermal) depends
entirely on the condensate assumptions.

Because we define the planet limb at the cloud tops,
the cloud wavelength-dependent albedos are not directly
important. Furthermore, even if the clouds were not optically thick,
they would not superimpose any spectral
features on the optical stellar flux since
their extinction (absorption + scattering) is generally a
smooth function of wavelength.

\subsubsection{Transparent Atmosphere Temperature-Pressure Profile}
\label{sec-TP}
We use the radial T-P profile
generated in the self-consistent
irradiated atmosphere code,
and construct a limb depth and line-of-sight T-P profile
from geometrical considerations.
The limb of the planet, as defined above, is approximately 0.01~$R_P$.
Together with $R_P=1.54 R_J$ and $R_*=1.3R_{\odot}$, this thickness
gives a ratio of
the projected transparent atmosphere to the stellar disk of
$3.6 \times 10^{-4}$.
The line-of-sight T-P profile
describes a plane-parallel
column of gas through which the stellar
flux passes. The gas is
optically thin; flux at most wavelengths passes through relatively
unattenuated. We compute the sum through several line-of-sight
columns of gas, from the densest column tangential to the cloud top
to the column that just skims the uppermost atmosphere.
However, in an atmosphere with exponential density fall off,
almost all of the absorption occurs in the densest column.

To solve for the radial
atmosphere structure we must make a specific assumption
about cloud particle type and size. Here we consider 10~$\mu$m grains
of MgSiO$_3$, Fe, and Al$_2$O$_3$.
For this particular model ($T_{\rm eff}=1350$~K), the transparent atmosphere above the limb
is comprised of a 
gas between 850-1000~K with pressures 20-600 dyne~cm$^{-2}$.
The main constituents of a gas at these temperatures and pressures
are H$_2$, CO, H$_2$O, and He. Because of the irradiative heating,
CO dominates
over CH$_4$ in this CEGP model.
Most gases are in molecular form with the exception
of He and the alkali metals.
The volatile elements (e.g. Mg, Ca, Ti), have condensed into grains
and we assume they have settled to within 1 pressure scale height
of the cloud base. Photochemistry
is not included, but the UV radiation from
the parent star could photoionize a small fraction of
H$_2$, CO, etc. through the transparent atmosphere. This, however,
would have little consequence on the results
(See \S\ref{sec-UVIR}).

\subsubsection{Cloud Models}
We caution that the transmission spectra presented here
are estimates that depend on several assumptions. Once
observations are successful, a more careful computation to accurately
interpret the data is necessary. 
One point to make is that the location of the cloud base ---
and hence the cloud tops --- depends on the irradiative heating
of the planet atmosphere,
which itself depends heavily on the absorbtivity of
the type and size of condensates
present throughout the planet atmosphere. A model
with absorptive grains causes the upper atmosphere temperature
to be higher compared to a model with highly reflective grains,
and the grain condensation boundary will be closer to the
top of the planet atmosphere. In the model used here
we consider 10~$\mu$m grains of MgSiO$_3$, Fe, and Al$_2$O$_3$.
While the MgSiO$_3$ clouds are the highest in the atmosphere,
all are important for computing the irradiated T-P
profile. The choice of cloud particle type and size distribution
is highly uncertain; this is the most complex free parameter of the
models.
In Seager et al. (1999) we discuss this in more detail.

Another major assumption used is that the radial T-P
profiles are similar on all parts of the planet. This would be the
case if strong winds redistribute the heat efficiently (e.g. as on
Jupiter where there is no apparent difference from one side of the
terminator to the other.) A different line-of-sight T-P
profile does not change our
predictions much; the flux-ratio is set by the planet limb, star area, and
planet area,
and the features are superimposed on that.
Indeed it is very difficult to predict the exact line shape
with so many unknowns about the planet atmosphere such as element
abundances (from both planet metallicity and non-equilibrium chemistry)
and cloud location due to heating.

\subsection{Results}
The CEGP and parent solar-type star have almost no spectral features
in common; at effective temperatures ($T_{\rm eff}$s)
of $\sim$1100~K to $\sim$1600~K, the CEGPs are almost 5 times
cooler than the stars at $T_{\rm eff}$'s 5000--6000~K. Thus, the
transmission spectra method is promising.
In this
subsection we discuss strong signals in the planetary
atmosphere.

\subsubsection{Alkali Metals}
\label{sec-alkali}
The Na~I and K~I resonance doublets are
predicted to be strong in CEGPs (Seager et al. 1999;
Sudarsky et al. 1999), assuming the atmospheres are similar
to brown dwarfs and cool L dwarfs which have similar $T_{\rm eff}$s.
Alkali metals have clear signatures in cool L dwarfs and brown dwarfs
where the metals such as Ti have condensed out of molecular form
into solids, depleting the strong optical molecular absorbers.
The K~I doublet $4^2S$ -- $4^2P$ absorption line
at 767.0~nm is extremely broad in methane dwarfs, such as Gliese 229B,
which have
$T_{\rm eff}$s a few hundred degrees lower than the CEGPs.
The broad wings of the K~I resonance doublet
extend for several tens of nm and are responsible for the large 
continuum depression in the optical and flux slope
redward to 1~$\mu$m (Tsuji et al. 1999; Burrows et al. 1999).
The broad lines are caused by strong pressure broadening by H$_2$,
and are in part so prominent because
there are no other strong absorbers present at those wavelengths
throughout the entire atmosphere.
There is some question about the shape of the alkali
metal lines in the CEGP atmospheres; they will be very broad if
the clouds are low in the atmosphere and the large pressures deep
in the atmosphere can contribute
to strong line broadening. Sudarsky et al. predict this scenario,
where K~I and Na~I absorb essentially all incoming optical radiation
redward of 500~nm. Sudarsky et al. use ad hoc modified T-P
profiles, to simulate heating, which have clouds at the 10 bar level.
Seager et al. (1999) include heating from
the parent star on the T-P profile,
and find that the lines are much narrower, since
clouds exist higher in the atmosphere at lower pressure.
Observations should be able to distinguish between the two cases.
%Nevertheless, in the transparent atmosphere
%of the planet, the K~I and Na~I resonance lines will be narrow --- 
%the deep pressure zones where the broad line wings are formed are
%not sampled.

With the columns of gas defined above (\S\ref{sec-TP}) we compute the attenuation
of the incoming intensity with the radiative transfer equation
in the limit
of no scattering $I(\nu,z)=I_*(\nu,z){{\rm exp} (- \kappa(\nu,z))}$,
where $I_*(\nu,z)$ is the stellar intensity, $\nu$ is the frequency,
and $z$ is the depth along the line-of-sight through the planet's
transparent atmosphere.
Here $\kappa(\nu,z)$ is the extinction which includes
absorption and scattering. 
Our code includes the dominant
opacities expected for cool L dwarfs and brown dwarfs:
H$_2$O, TiO, CH$_4$, H$_2$-H$_2$ and H$_2$-He collision induced opacities,
Rayleigh scattering from H, He, H$_2$,
% (using the Rayleigh scattering phase function),
and the alkali metal lines.
The oscillator strengths
and energy levels for the lower levels of the alkali metals (Na, K, Li, Rb, Cs)
were taken from Radzig \& Smirnov (1985) and the Kurucz
atomic line list (Kurucz CD ROM 23), and  
we compute line broadening using a Voigt profile with H$_2$ and He
broadening and Doppler broadening.
We do not need to solve the full radiative
transfer equation since we assume the effect of transmitted intensity
through the planet's transparent atmosphere
is negligible with regard to the radiative structure which is
already accounted for in the irradiation model.

Figure~1 shows the flux from the star and the stellar flux
that has passed through the planet's transparent atmosphere.
The curves are
essentially the same in the UV through the optical, with the exception
of the absorption lines, including Na~I at 285.4~nm and 342.8~nm,
the Na~I resonance doublet at 589.4 nm and the K~I resonance
doublet at 767.0~nm. The He~I triplet line
is at 1083~nm. In the infrared, the stellar flux is absorbed
by the water bands, as well as the 5$\mu$m and the 3.3$\mu$m
methane band (not visible in the figures). No molecular features such as TiO appear
in the spectra, since those molecules have been depleted into solids.
Solar-type stars also have alkali
metal lines but they are weak because most of the alkali metals
are ionized.
In addition, the stellar optical spectrum is crowded with other atomic
absorption lines. 

Figure 2 shows the normalized in-transit minus
out-of-transit spectra, i.e. the transmission spectra as
the percentage occulted area 
of the star at different wavelengths. In transmitted
light the planet is different sizes at different wavelengths.
The zero point (at $\sim$ 1.47\%) is set by the atmosphere depth
at which the planet is optically thick at all wavelengths (the
observed transit).
The planet appears largest in the line cores, where photons passing
very high in the planet atmosphere are still absorbed along
the line of sight due to the strong absorption by the
Na~I and K~I resonance lines. This figure differs from the model
in Figure~1 in that we have considered the atmosphere out
to the distance where the Na~I resonance line becomes optically
thin along the line of sight. In this case the limb depth
is several percent
of $R_P$ (where the limb coincides with the observed transit).
Rayleigh scattering from H$_2$ is important below 200~nm;
otherwise it is negligible.
The Na~I resonance doublet is broader than the K~I resonance
doublet because its abundance is an order of magnitude higher.

Figure~2 also shows the transmission spectra from
an atmosphere with a cloud base much deeper
than predicted by our model, at 0.2 bar instead of at
2.4$\times10^{-3}$~bar.
(We note that in our range of models for different condensate type
and size the highest clouds have bases
ranging from roughly 0.5 to $10^{-3}$~bar.)
There will be two main consequences.
First the transparent atmosphere area will be larger, resulting
in the planet's total line
depths being larger with respect to the zero point.
Second, the stellar flux will pass through higher
densities, pressures, and temperatures.
Rayleigh scattering is strong
in this relatively high density transparent atmosphere.
When higher pressures are sampled by the transmitting photons,
the lines become more pressure broadened.
This is seen in the Na~I and K~I resonance doublets.
The higher densities will cause stronger alkali lines and additional
absorption lines
from other transitions that were too weak to appear in the low densities
of our line-of-sight model, for example, non-resonance lines
of Na~I and K~I.
Observations of the alkali metal lines will be able to constrain cloud location.
Exospheric escape or photoionization by stellar UV radiation could
affect the line cores.

\subsubsection{Neutral Helium}
We expect a strong absorption line at 1083~nm due to scattering of
background stellar photosphere photons off the overpopulated He atoms
excited to the triplet $2^3S$ metastable state. The mechanism works
as follows. The stellar EUV radiation shortward of 50.4~nm 
will photoionize neutral He atoms
and they will recombine at the local kinetic temperature, which may be
as low as 800~K in the upper atmosphere of the planet. The He~I recombination
cascade is efficient for the singlet states, but stops at $2^3S$ for the
triplet states which lack a fast radiative decay path to the ground state.
Due to the low local kinetic temperature, collisional de-excitation is
negligible. On the other hand, the number of 1083~nm continuum
photons from the G0V star is very large $-$ they scatter efficiently
in the $2^3S-2^3P$ transition and produce the strong absorption
feature in the transmission spectrum.
 
Versions of this mechanism are responsible for enhanced He~I 1083~nm 
absorption in solar spectra (Zirin 1975; Andretta \& Jones 1997),
Algol binary systems (Zirin \& Liggett 1982), etc.
In the case of HD~209458, a sun-like star, we assume the EUV radiation
to be that of the Sun (from Tobiska 1991). We use the T-P distribution of the
transparent atmosphere generated in the Seager \& Sasselov code and illuminate it with
that diluted EUV field, simultaneously solving the NLTE transfer for a
helium model atom with singlet and triplet states up to $n=4$. The details
of this calculation are essentially the same as in Sasselov \& Lester (1994).
One has to realize that this calculation is more of an estimate than an
fully consistent treatment of He~I 1083~nm line formation in the unusual
conditions of the EGP's atmosphere. However, the mechanism described above
is very robust and given no other competing target(s) for the EUV photons except
H and H$_2$, the $2^3S-2^3P$ transition is optically thick at line center.
In fact, He (and H) are prone to creating an extended exosphere around
the planet; if so, the strength of the He~I 1083.0~nm absorption may well
be extremely strong (due to a much larger transparent atmosphere)
and easy to observe.
The broadening of the line ($\sim$~0.3~nm) is not significant, but this 
needs further study in terms of He collisional broadening by molecular
species like H$_2$. 

In the spectrum of the inactive parent star, the He~I 1083.0~nm triplet line
is extremely weak, if it is present at all. This makes it a promising signature
in the combined planet-star flux.
Enhanced absorption of the $3^3P-3^3D$ transition
of C is also seen in binary systems,
but in the CEGP atmospheres the C is locked in CO or CH$_4$.

\subsubsection{UV and Infrared Wavelengths}
\label{sec-UVIR}
Solar System outer planet occultation transmission lines have
been very successfully observed in the UV where
the molecules such as H$_2$, N$_2$, and O$_2$ have strong absorption signatures.
In addition, in CEGPs the H resonance line transitions are strong
(and may be enhanced by photodissociation of H$_2$)
and alkali metal resonance line absorption
appears in the near-UV (Figure~2).
For the CEGPs orbiting sun-like stars, the UV flux is very low, and the
lines --- while useful diagnostics --- will be difficult to detect.

The spectral transmission signatures in the infrared
(e.g. H$_2$O and CH$_4$)
shown in Figure~1 will be difficult to distinguish from the planet's own
thermal emission, which is present at all phases.
More importantly, redward of 2000~nm,
the planet's thermal emission which roughly follows a blackbody
(dotted line in Figure~1), will be much stronger
than the transmitted spectrum which follows the optical-peaking
stellar spectrum.
This is in contrast to the optical where the planet has no emission;
during transit the optical transmission spectrum is the planet's only
contribution to the total flux.
Blueward of 2000~nm the transmission spectrum may be brighter
than the planet's own emission, and for the $2.4 \times 10^{-3}$ bar
cloud base model, observations in this region may be very promising.
(Note that for some models 
the planet's thermal emission can be much
higher than a blackbody in that region (Seager 1999)).
%the transmission spectrum actually dominates over the planet's
%own emission. Nevertheless,
CH$_4$ is an important temperature diagnostic, because
the temperature-pressure profiles of the CEGPs fall near the CO/CH$_4$
equilibrium curve (Seager 1999, Goukenleuque et al. 1999). The strength
of the methane absorption could indicate the temperature in the
planet's upper atmosphere layers, which in turn will distinguish between
different irradiated models.

In principle the infrared magnitude of the brightness of the
planet will increase slightly just after first contact as illumination passes
through the top of the transparent atmosphere, but against the backdrop
of the star this will not be detectable. 
CO (not included in this model) may also be 
good transmission signatures in the infrared.

\subsubsection{Comparison of spectra using different $R_P$ and R$_*$}
We have also run calculations using $R_P=1.27$ (and $i=87^{\circ}$)
from Charbonneau et al. (1999a), who assumed $R_*=1.1$.
In this case the ratio of planet-to-star area is slightly
lower ($\sim$ 5\%) than with the correct values due to
the higher inclination.
In our self-consistent models,
the effect of a smaller planet radius and smaller
stellar radius decreases the
transmitted-to-stellar flux ratio by an even smaller amount. The reason
is that a smaller star and planet means the star-planet distance
is larger, making the planet's atmosphere slightly cooler. The cooler
atmosphere has the MgSiO$_3$ cloud-top lower in the atmosphere
and hence has a larger transparent atmosphere area compared
to the hotter model. For example, the equilibrium effective temperature
is $T_{\rm eq} = T_* (R_*/2D)^{1/2} [(1-A)]^{1/4}$, where $T_*$ is the effective
temperature of the star, and $A$ is the Bond albedo.
Because $T_{\rm eq} \sim R_*^{1/2}$, decreasing
$R_*$ from 1.3$R_{\odot}$ to 1.1$R_{\odot}$ will change the planet's
$T_{\rm eq}$ by $\sim$100~K.

There is also a small change due to density. At the same optical
depth, the smaller radius planet with a higher surface gravity
will have a higher density compared to the planet
with a larger radius. Thus in comparison the absorption
lines will be slightly stronger, for the same optical depth.

\section{Other EGPs}

As of this writing there are 29 known extrasolar planets around 27 stars.
$M \sin i$ ranges from 0.42 to 11 $M_J$ and planet-star distances from
0.042~AU to 2.5~AU. In principle observations of the transmission features
of any transiting planet can be attempted. 
Many of the stars are being monitored for transits,
and transits around several of these stars have been excluded
(Henry et al. 2000, 1997; Baliunas et al. 1997).
Three others of these known planet systems have their inclinations
limited by observation of
Kuiper-belt-like disks (Trilling et al. 1999).
Wide-field transit searches are
excluded to transiting systems and are mostly sensitive to systems
with orbital distances below 0.2~AU.

CEGPs much more massive than HD~209458~b will have a much smaller
transparent atmosphere area. 
The planet radius is a weak
function of mass (Guillot et al. 1996) and more massive
planets are expected to be more compact as the degenerate core
increases at the expense of the gaseous atmosphere. For a more massive
planet with all other parameters
equal (atmosphere structure, radius, etc.), the scale height is smaller,
and defined by optical depth the entire atmosphere
including the transparent atmosphere is smaller.
For example $\tau$~Boo~b which has
$M \sin i=3.87$ (Butler et al. 1997) is at least 5.6 times more massive
than HD~209458~b.
CEGPs with higher $T_{\rm eff}s$ than HD~209458~b (e.g $\tau$ Boo~b)
may have cloud bases
closer to the top of the atmosphere, also resulting in a smaller transparent
atmosphere area.
The flux ratio of transmitted to total stellar flux is also
sensitive to the stellar radius, and would change by a factor
of three for solar-type parent stars;
from evolutionary calculations
(Ford, Rasio, \& Sills 1999) the solar-type parent stars of known CEGPs
with orbital distances below $\sim$0.2~AU range
 from $0.93$--$1.56~R_{\odot}$.

Wide-field transit searches will find short-period planets orbiting
stars where radial velocity planet detections are not possible, for example
around active cool stars and hot stars that have rotationally
broadened atomic lines and activity.
The parent stars of known EGPs
range from F6IV ($\tau$ Boo) down to M4V (Gliese 876).
Planet transmission spectra may be difficult to disentangle
from parent M stars whose optical spectra are very crowded with
molecular lines such as TiO and VO.
Nevertheless, because of the larger planet-to-star
area ratio, the flux ratio is enhanced by a factor of $\sim$10.
Jupiter-sized planets orbiting hot stars such as a B or O star
would have a flux ratio decreased compared to solar by
a factor of 10-100, but the UV
flux (observed from space) would provide many useful molecular
absorption signatures such as H$_2$, N$_2$, and O$_2$.
%, photoevaporation of disks may mean no planets.

\section{Summary and Prospects}
We have estimated the transmission spectra
of a CEGP during an occultation of the parent star.
We find very strong absorption signatures of Na~I and K~I,
and a strong signature of the He~I $2^3S$ -- $2^3P$
triplet line at 1083.0~nm.
 We find the number, strength, and
depth of spectral features are
sensitive to the cloud-top depth in the planet atmosphere.

%absorption caused by photoionization
%of He~I followed by recombination to the .
Detecting spectral features will require
high resolution, high signal-to-noise observations (e.g. with
Keck HIRES).
During the transit of HD~209458, the Doppler shift of the planet
is strong enough so that it should be taken into account when
analysing the spectra. Spectral separation techniques, designed
to detect absorption signatures at the $< 10^{-4}$ level
(e.g. Cameron et al. 1999; Charbonneau et al. 1999b),
may be necessary to recover the weak planet signal.
Systematic red or blue shifting from winds blowing between
the day and night side may have a detectable effect on the planet spectrum
(D. Charbonneau and T. Brown, private communication). 
Measurements of metallic lines will also constrain (together with $M_P$,
$R_P$, and ${\rho}_P$) the interior models and shed light on the formation
scenario of CEGPs, e.g. core accretion vs. gravitational disk instability.
The CEGP's extended exosphere may be easy to detect in the He~1083.0~nm transition.

If successful,
observations of transmission spectra will be the first made of an extrasolar
planet atmosphere,
and will provide constraints on the upper atmosphere column density, temperature,
and pressure. In addition the observations should easily constrain
the cloud-top depth, which naturally defines the planet limb.
This information will help distinguish between atmosphere models.
Most importantly, detection of the alkali metal absorption lines will confirm
the very basic postulate that the CEGPs have similar atmospheres
to those of methane
dwarfs and cool L dwarfs which have similar $T_{\rm eff}$s.

\acknowledgements{ We are grateful to Dave Latham for providing
the stellar and planet parameters for the HD~209458 system before
publication.
We thank Bob Noyes, Tim Brown and Mark Marley
for reading the manuscript and for helpful comments and discussion.
We also thank Dave Charbonneau and Avi Loeb for useful discussions.
SS is supported by NSF grant PHY-9513835. DDS acknowledges support from
the Alfred P. Sloan foundation.}

 Note added in proof: Mazeh et al. (2000) gives
$R_P = 1.4 \pm 0.17 R_J$ and $R_*=1.2 \pm 0.1 R_{\odot}$,
corrected from an earlier version quoted in this paper.
We have not incorporated these values in this paper
which is meant to be a conceptual description and
estimate of CEGP transmission spectra rather
than an exact prediction.

\begin{figure}
\plotone{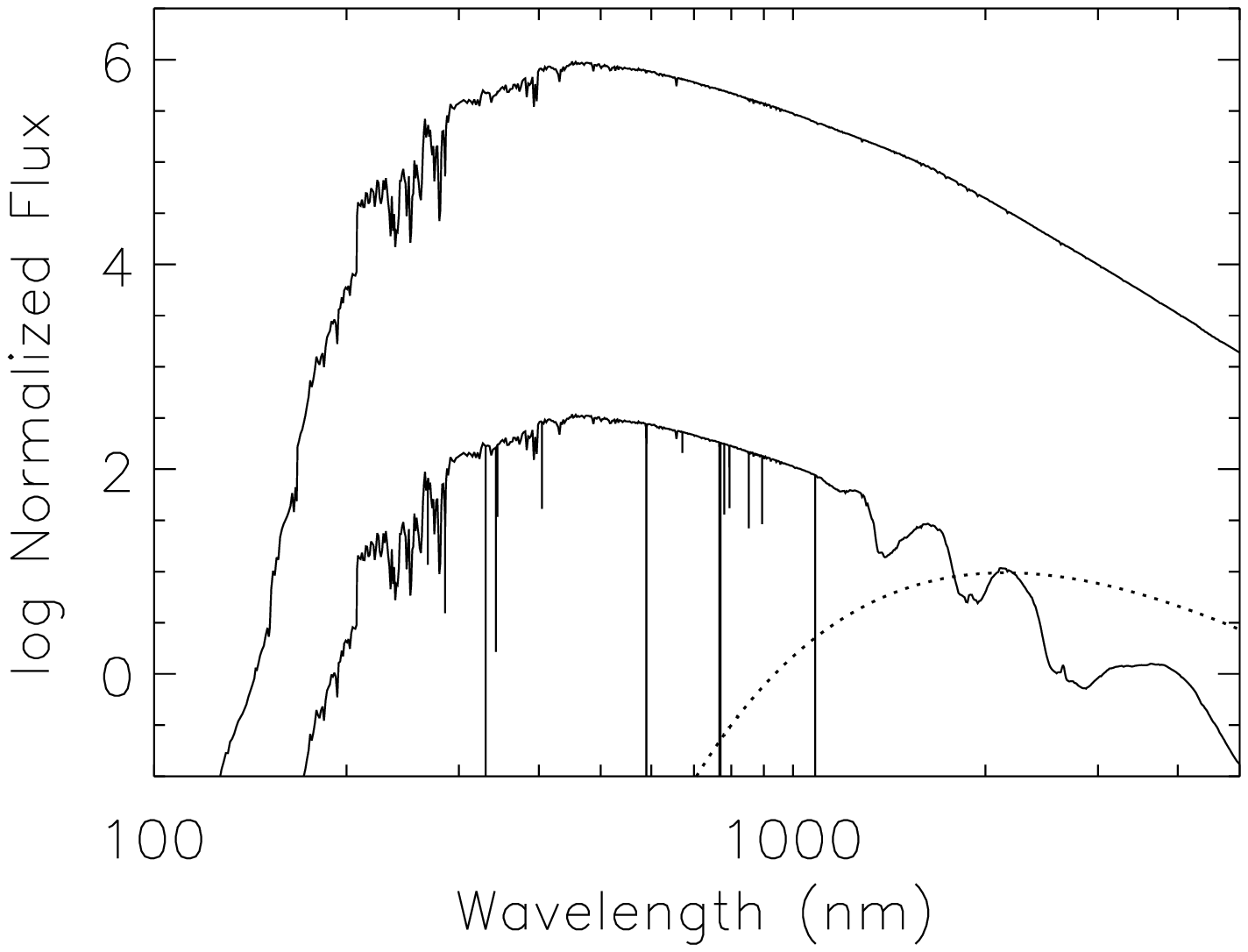}
\caption{The flux of HD~209458~A (upper curve) and the
transmitted flux through the planet's transparent atmosphere (lower curve).
Superimposed on the transmitted flux are the planetary absorption
features, including the He I
triplet line at 1083 nm. The other bound-bound lines are
alkali metal lines (see Figure~2 for details).
The H$_2$O and CH$_4$ molecular absorption dominates in
the infrared. The dotted line
is a blackbody of 1350~K representative of the
CEGP's thermal emission, but the thermal emission can be larger
than a blackbody blueward of 2000~nm.}
\end{figure}

\begin{figure}
\plotone{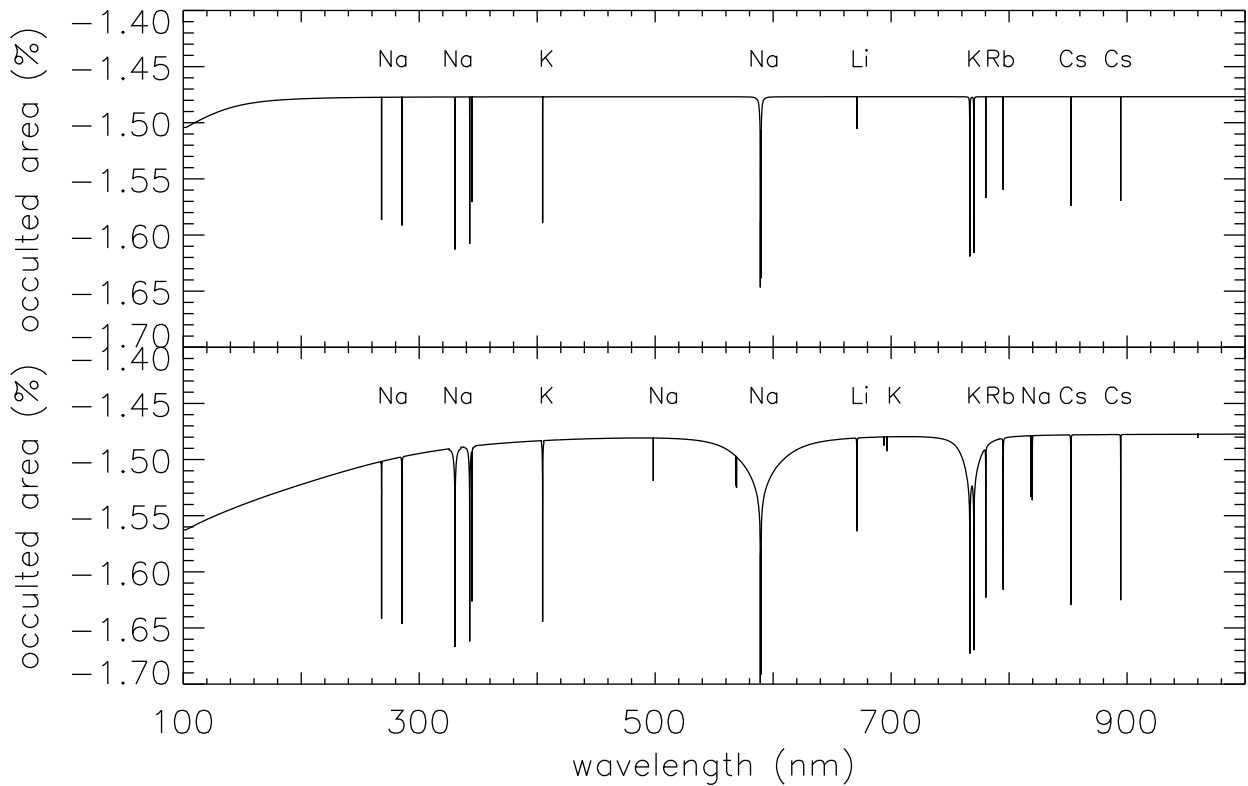}
\caption{Upper plot: the normalized in-transit minus
out-of-transit spectra, i.e. percent occulted area of the star.
In this model the cloud base is at 2.4$\times10^{-3}$ bar.
Rayleigh scattering is important in the UV.
Lower plot:
a model with cloud base at 0.2 bar. The stellar flux passes
through higher pressures, densities, and
temperatures of the planet atmosphere compared to the model
in the upper plot. In addition, a larger transparent
atmosphere makes the line depth larger.
Observations will constrain the cloud depth.
See text for discussion.
% resulting in stronger and broader
%lines, additional absorption lines from more absorbers, and more H$_2$
%Rayleigh scattering. Observations
%should be able to constrain the cloud location.
}
\end{figure}

\end{document}